\documentclass{article}

\usepackage{arxiv}

\usepackage[utf8]{inputenc} 
\usepackage[T1]{fontenc}    
\usepackage{hyperref}       
\usepackage{url}            
\usepackage{booktabs}       
\usepackage{amsfonts}       
\usepackage{nicefrac}       
\usepackage{microtype}      
\usepackage{lipsum}		
\usepackage{graphicx}
\usepackage[square,sort,comma,numbers]{natbib}
\usepackage{doi}
\usepackage{multirow}
\usepackage{float}
\usepackage{booktabs, makecell, tabularx}
\usepackage{graphicx}
\usepackage{booktabs}
\usepackage{amsmath}
\usepackage{multirow}
\usepackage{float}
\usepackage{graphicx}
\usepackage{xcolor}
\usepackage{hyperref}
\usepackage{multirow}
\usepackage{rotating}
\usepackage{pifont}
\newcommand{\cmark}{\ding{51}}%
\newcommand{\xmark}{\ding{55}}

\title{Rethinking Attention Gated with Hybrid Dual Pyramid Transformer-CNN for Generalized Segmentation in Medical Imaging}

\author{ \href{https://sciprofiles.com/profile/FaresBougourzi}{\includegraphics[scale=0.06]{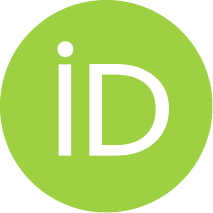}\hspace{1mm}Fares BOUGOURZI}
\thanks{.} \\
Junia, UMR 8520, CNRS, Centrale Lille, Univerity of Polytechnique Hauts-de-France, 59000 Lille, France \\
	\texttt{faresbougourzi@gmail.com; fares.bougourzi@junia.com} \\
	\And
	\href{https://orcid.org/0000-0001-5077-4862}{\includegraphics[scale=0.06]{orcid.pdf}\hspace{1mm}Fadi DORNAIKA} \\
	University of the Basque Country UPV/EHU,\\
	San Sebastian, SPAIN; IKERBASQUE, Basque \\ Foundation for Science, Bilbao, SPAIN \\
	\texttt{fadi.dornaika@ehu.eus} \\ 
	\And
	\href{https://orcid.org/0000-0000-0000-0000}{\includegraphics[scale=0.06]{orcid.pdf}\hspace{1mm}Abdelmalik Taleb-Ahmed} \\
	Universit{\'e} Polytechnique Hauts-de-France, Université de Lille, \\CNRS, Valenciennes, 59313, Hauts-de-France, France\\
	\texttt{Abdelmalik.Taleb-Ahmed@uphf.fr} \\
	\And
	\href{https://orcid.org/0000-0000-0000-0000}{\includegraphics[scale=0.06]{orcid.pdf}\hspace{1mm}Vinh Truong Hoang} \\
	Ho Chi Minh City Open University, Viet Nam,\\
	\texttt{vinh.th@ou.edu.vn} \\ 
}

\renewcommand{\shorttitle}{\textit{arXiv} Template}

\hypersetup{
pdftitle={A template for the arxiv style},
pdfsubject={q-bio.NC, q-bio.QM},
pdfauthor={David S.~Hippocampus, Elias D.~Striatum},
pdfkeywords={First keyword, Second keyword, More},
}

\begin{document}
\maketitle

\begin{abstract}

Inspired by the success of Transformers in Computer vision, Transformers have been widely investigated for medical imaging segmentation. However, most of Transformer architecture are using the recent transformer architectures as encoder or as parallel encoder with the CNN encoder. In this paper, we introduce a novel hybrid CNN-Transformer segmentation architecture (PAG-TransYnet) designed for efficiently building a strong CNN-Transformer encoder. Our approach exploits attention gates within a Dual Pyramid hybrid encoder. The contributions of this methodology can be summarized into three key aspects: (i) the utilization of Pyramid input for highlighting the prominent features at different scales, (ii) the incorporation of a PVT transformer to capture long-range dependencies across various resolutions, and (iii) the implementation of a Dual-Attention Gate mechanism for effectively fusing prominent features from both CNN and Transformer branches. Through comprehensive evaluation across different segmentation tasks including: abdominal multi-organs segmentation, infection segmentation (Covid-19 and Bone Metastasis), microscopic tissues segmentation (Gland and Nucleus). The proposed approach demonstrates state-of-the-art performance and exhibits remarkable generalization capabilities. This research represents a significant advancement towards addressing the pressing need for efficient and adaptable segmentation solutions in medical imaging applications.

\end{abstract}

\keywords{Transformer \and Convolutional Neural Network \and Deep Learning \and Medical Imaging \and Segmentation \and Unet \and Synapse \and (Gland and Nucleus) \and Covid-19.}

\section{Introduction}
Medical imaging segmentation plays a crucial role in diagnosing, assessing severity, and monitoring progress in various medical conditions \cite{shamshad_transformers_2022}. Despite significant advancements in utilizing machine learning for medical imaging segmentation, several challenges persist in developing efficient segmentation approaches. These challenges include limited labeled data availability, which is a laborious and error-prone task \cite{shamshad_transformers_2022,bougourzi_emb-trattunet_2024}. The ultimate goal remains to devise a generalized approach for different medical segmentation tasks. However, achieving efficiency across various medical imaging segmentation tasks remains challenging due to the high variability among diseases, ranging from single classes to multi-classes, and from disease to organ segmentation. Consequently, many approaches are tailored to specific tasks, limiting their applicability to other tasks.

In the last decade, Convolutional Neural Networks (CNNs) have emerged as the primary approach for medical imaging segmentation \cite{shamshad_transformers_2022, bougourzi_emb-trattunet_2024, khan_transformers_2021, chen2021transunet}. However, CNNs are predominantly adept at extracting local features, thereby overlooking long-range dependencies, which are crucial for modeling global contextual features. Transformers have demonstrated high capability in encoding long-range dependencies, leading to their integration into segmentation architectures either as pure architectures or hybrid ones combined with CNNs \cite{shamshad_transformers_2022, wang2022uctransnet, khan_transformers_2021, chen2021transunet, wang_transbts_2021}. However, existing architectures often utilize transformers as single or parallel encoders alongside CNN encoders \cite{chen2021transunet, wang_transbts_2021, hatamizadeh_unetr_2022, zhu2023brain, wu_fat-net_2022, he2023medical, wang2022mixed}, indicating limitations in efficiently combining transformer and CNN features.

To address this, we propose revisiting attention gates to build a stronger encoder, introducing our Dual-Attention Gate. Unlike conventional attention gates originally designed to select prominent features from the encoder during decoding \cite{oktay_attention_2018}, our Dual-Attention Gate selects prominent features between CNN features via an input pyramid and from the transformer branch via the main CNN feature path. This results in a more compact main path.

The paper introduces a novel approach called PAG-TransYnet, which combines Transformer and CNN architectures using Dual-Attention Gates. These gates aim to extract significant feature regions and merge features from both CNN and Transformer models. The encoder structure of PAG-TransYnet consists of three branches. The first branch undergoes contraction through four pyramid levels using convolutional blocks, producing features that act as a gating signal for highlighting prominent features in the second branch. The second branch, termed the main branch, focuses on extracting features from the input data. Simultaneously, the features from the main branch are used to highlight important features in the third branch, which utilizes Transformer architecture. The attention features from both branches are concatenated to form the new main branch features for the subsequent level. Overall, the proposed approach aims to capture both local and global features through attention mechanisms, resulting in a comprehensive representation of the input data.

\begin{figure}
\setlength\tabcolsep{10pt}
\begin{tabularx}{\linewidth}{XXXXXp{1pt}}
\\
    \includegraphics[width=3cm, height=3cm]{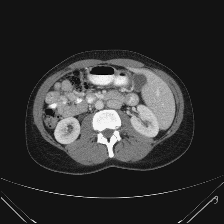} &     
     \includegraphics[width=3cm, height=3cm]{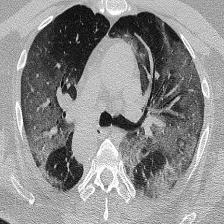} &

     \includegraphics[width=3cm, height=3cm]{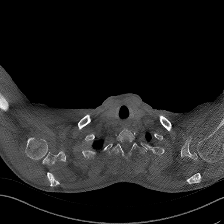} & 

     \includegraphics[width=3cm, height=3cm]{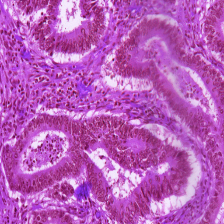} &
     \includegraphics[width=3cm, height=3cm]{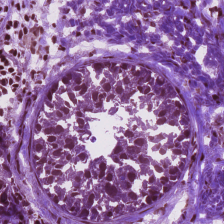} &
\\

     \includegraphics[width=3cm, height=3cm]{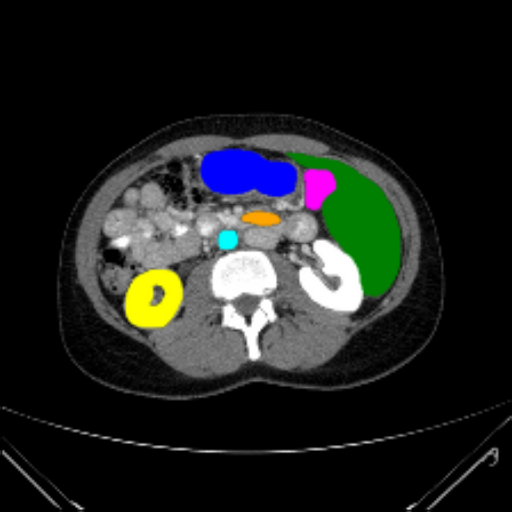} &     
     \includegraphics[width=3cm, height=3cm]{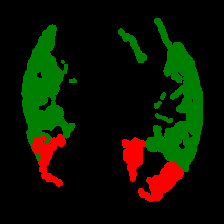} &

     \includegraphics[width=3cm, height=3cm]{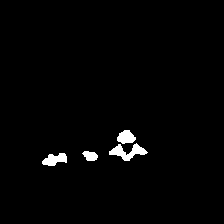} & 

     \includegraphics[width=3cm, height=3cm]{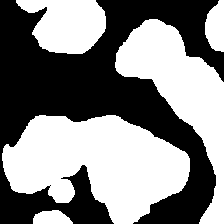} &
     \includegraphics[width=3cm, height=3cm]{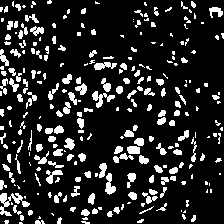} &
     
\\
     \includegraphics[width=3cm, height=3cm]{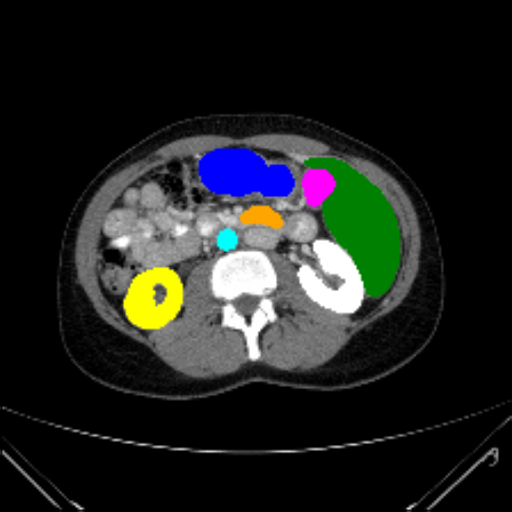} &     
     \includegraphics[width=3cm, height=3cm]{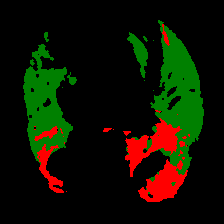} &

     \includegraphics[width=3cm, height=3cm]{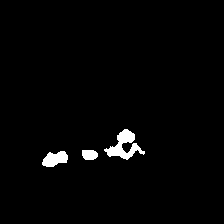} & 

     \includegraphics[width=3cm, height=3cm]{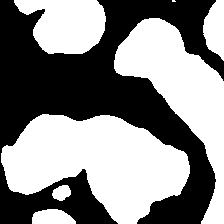} &
     \includegraphics[width=3cm, height=3cm]{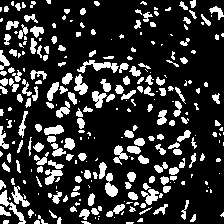} &     
\end{tabularx}
\caption{Examples of Medical Imaging Segmentation, the first, second and third rows represent the input image, ground truth and the prediction of our approach, respectively. First, second, third, fourth and fifth columns depict abdominal multi-organ segmentation,  Covid-19, Bone Metastasis, Gland, and Nucleus, respectively.}
\label{fig:segexmp}
\end{figure}

In summary, the main contributions of this work are:
\begin{itemize}
\item   Introduction of a novel hybrid architecture for medical imaging segmentation, which seamlessly integrates CNN, Transformers, and a fusion branch encoder.

\item Enhancement of the Att-Unet attention gate through our proposed Dual-Attention Gate. This refinement involves redesigning its structure, repositioning it within the encoder, and optimizing its functionality within the fusion objective.

\item Demonstration of the remarkable capability of our approach to achieve state-of-the-art performance across a diverse range of medical imaging segmentation tasks, including organ scans segmentation, infection detection, and microscopic tissue segmentation (Fig. \ref{fig:segexmp} shows examples of the considred segmentation tasks).

\item   The code for PAGTransYnet are made publicly available at \url{https://github.com/faresbougourzi/PAGTransYnet}.

\end{itemize}

This paper is organised as follows: Section 2 highlights the related works. In section 3, the proposed approach is described. Section 4 depicts and analyzes the obtained results. Finally, section 5 concludes this paper.

\section{Related Works}

In recent years, Convolutional Neural Networks (CNNs) have achieved state-of-the-art performance in medical image segmentation, particularly following the proposition of the U-Net architecture by Ronneberger et al. in 2015 \cite{ronneberger_u-net_2015}. Since then, numerous variants such as Attention U-Net (Att-UNet) \cite{oktay_attention_2018}, U-Net++ \cite{zhou_unet_2018}, and ResU-Net \cite{zhang_road_2018} have emerged, each aiming to enhance segmentation performance. The U-Net architecture, characterized by an encoder-decoder structure with skip connections, has proven effective in preserving fine-grained details through feature concatenation. On the other hand, attention mechanisms have been widely investigated for medical imaging segmentation. One of the most famous attention mechanisms is the Attention Gate (AG), proposed by Oktay et al. in 2018 \cite{oktay_attention_2018}, which integrates attention into U-Net after the skip connection, producing a variant known as Att-UNet. The main objective of Att-UNet is to highlight salient regions in encoder features using decoder features.
However, the efficacy of attention gates can vary, prompting the introduction of our approach: the Dual-Attention Gate, integrated into the encoding phase, leveraging Pyramid features, CNN features, and Transformer features to enhance feature extraction and emphasize prominent regions.
Despite the great success of CNNs in medical imaging segmentation, their main shortcoming lies in their weakness in capturing long-range dependencies, as CNNs are primarily focused on extracting local features \cite{shamshad_transformers_2022, hatamizadeh_unetr_2022, zhu2023brain, bougourzi2023pdatt}. On the other hand, Transformers, renowned for their ability to capture long-range dependencies in sequences, have shown promising performance in medical imaging tasks, including classification, detection, and segmentation \cite{shamshad_transformers_2022, wu_fat-net_2022, he2023medical, wang2021transbts, wang2022mixed}. In segmentation, both 2D and 3D transformer-based approaches, such as Fat-Net and U-Transformer, have showed promising performance by fusing CNN and Transformer components to enhance segmentation accuracy \cite{wu_fat-net_2022, petit_u-net_2021, hatamizadeh_unetr_2022}.

The integration of CNN and Transformer blocks into single architectures has been a focal point, particularly in the encoding phase \cite{hatamizadeh_unetr_2022,zhu2023brain,wu_fat-net_2022,he2023medical,wang2021transbts,wang2022mixed}. Various encoder configurations have been proposed, including solely Transformer-based encoders \cite{hatamizadeh_unetr_2022,zhu2023brain}, parallel CNN and Transformer encoders with subsequent fusion \cite{wu_fat-net_2022,he2023medical}, and CNN encoders followed by Transformer blocks \cite{wang2021transbts,wang2022mixed}. However, many existing approaches lack robust connectivity between Transformer and CNN features, indicating a gap in feature integration. To address this, our approach introduces a novel encoder architecture incorporating Pyramid features, CNN features, Transformer features, and Dual-Attention Gates, aiming to significantly enhance feature fusion and improve segmentation performance.


\begin{figure}[ht]
\centering
    \includegraphics[width = 5in, height = 2in] {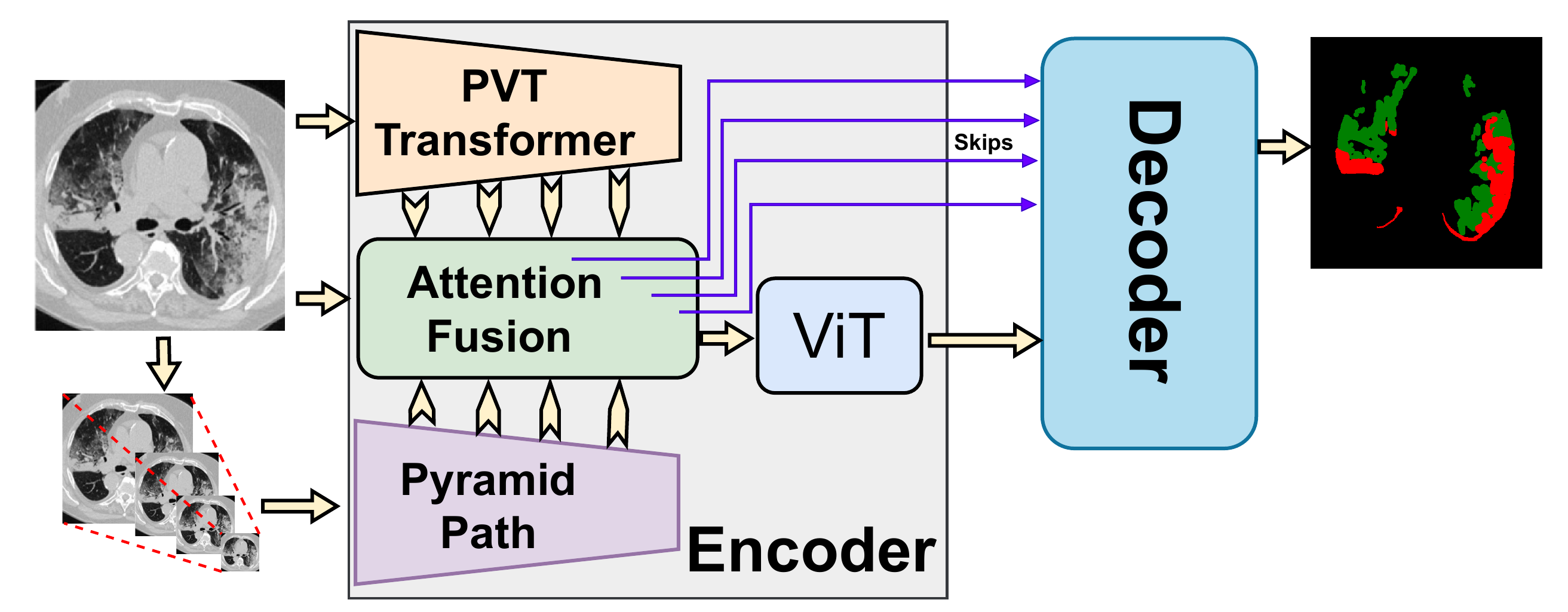} 
    \caption{Our proposed PAG-TransYnet architecture. }
    \label{fig:approach}
\end{figure}


\begin{figure}[ht]
  \centering
  \includegraphics[width = 7in, height = 3.7in]{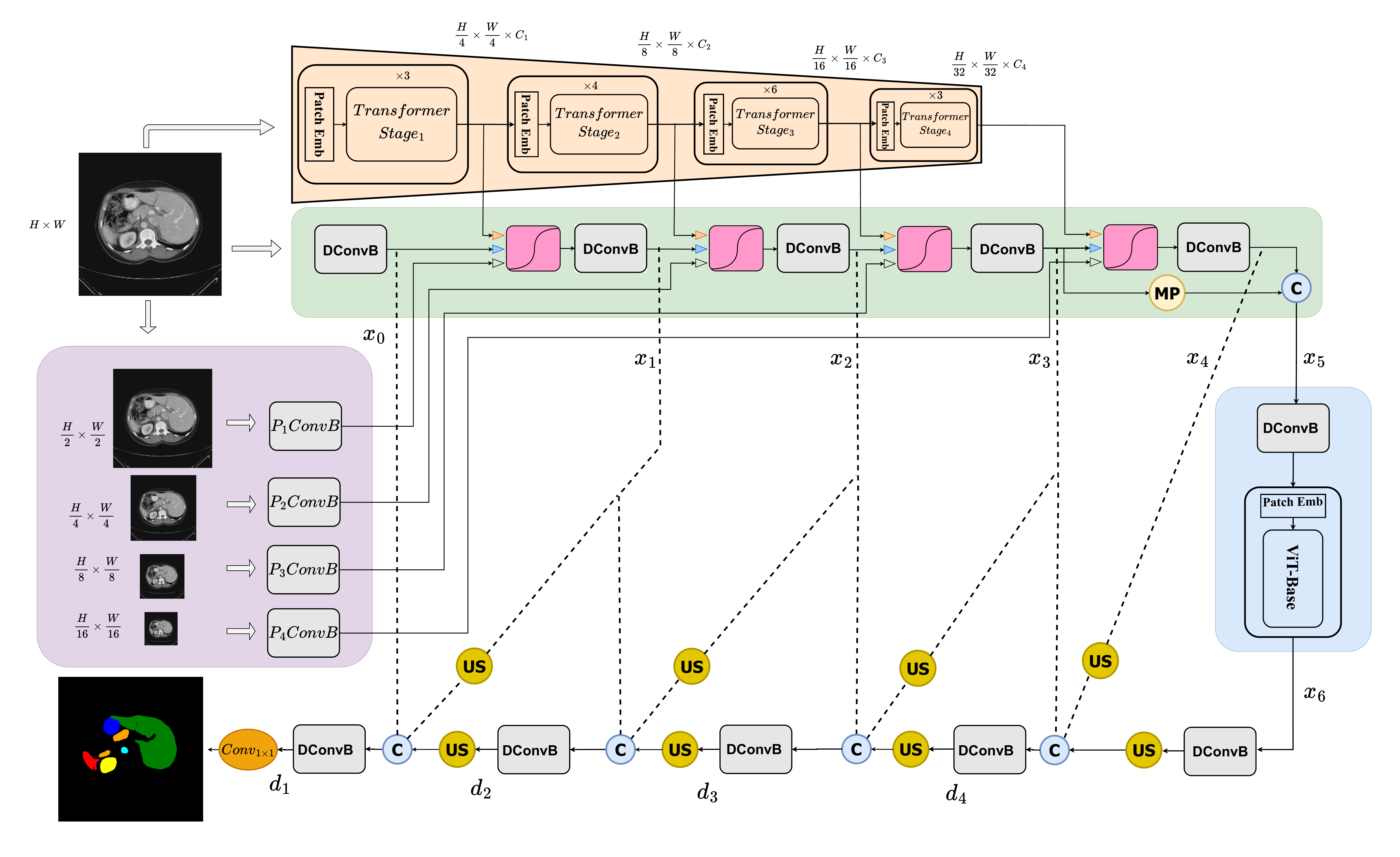} 
  \caption{The detailed description of our proposed PAG-TransYnet approach.}
  \label{fig:approach.2}
\end{figure}
\begin{figure}[ht]
  \centering
  \includegraphics[width=1\textheight,height=0.27\linewidth,keepaspectratio]{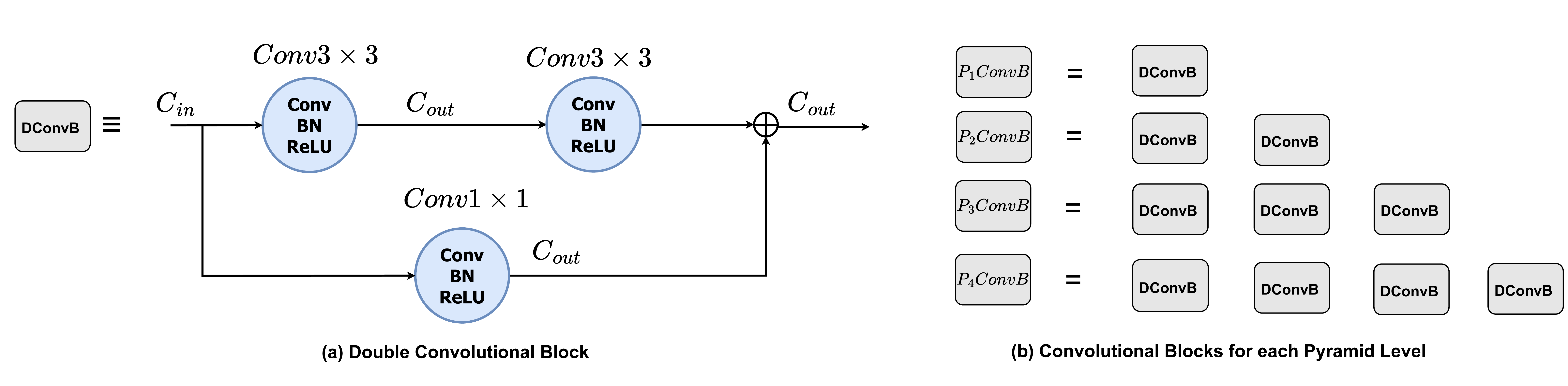} 
  \caption{Detailed representation of the convolutional blocks used in our proposed PAG-TransYnet architecture.}
  \label{fig:dconv}
\end{figure}
\begin{figure}[ht]
  \centering
  \includegraphics[width=1\textheight,height=0.3\linewidth,keepaspectratio]{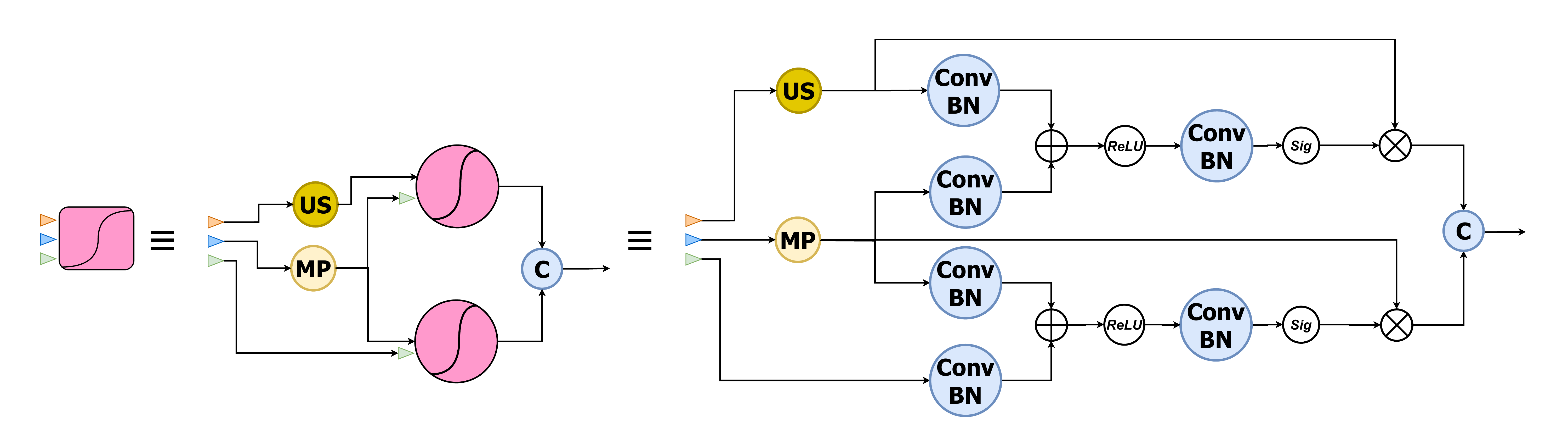} 
  \caption{The proposed Dual-Attention Gate, left the general structure and the right is the detailed one.}
  \label{fig:dag}
\end{figure}

\section{Proposed Approach}

Our proposed Pyramid Dual-Attention Gate Transformer-Ynet (PAG-TransYnet) has three encoder branches and a Unet-like decoder as shown in Figure \ref{fig:approach}. The detailed architecture is illustrated in Figure \ref{fig:approach.2}. 

As shown in Figure \ref{fig:approach},  the encoder of our proposed architecture consists of four main components: (i) a Pyramid Vision Transformer (PVT-v2), (ii) a pyramid representing the input image with four levels, each level followed by convolution blocks ($P_1ConvB$, $P_2ConvB$, $P_3ConvB$ and $P_4ConvB$), (iii) a main encoder path that  merges the PVT features and the main encoder features using Dual-Attention Gates, and (iv) a classic Transformer (Base ViT) serving as the final stage of encoding.

\subsection{Pyramid Encoder}

The pyramid encoder branch aims to provide convolutional features at four levels of the input image pyramid, which are subsequently utilized in the spatial gate attention mechanism. The image undergoes transformation into a pyramid with four levels, each level being resized separately. 
There are four pyramid levels, each with a pyramid input ($P_1$, $P_2$, $P_3$, and $P_4$) derived from the input image ($I$). These pyramid levels generate pyramid feature maps ($P_{f_1}$, $P_{f_2}$, $P_{f_3}$, and $P_{f_4}$) using pyramid convolutional blocks (PConvB), which consist of double convolutional blocks (DConvB). Notably, the first pyramid level contains one DConvB, whereas the fourth level incorporates a cascade of four DConvBs, as shown in Figure \ref{fig:dconv}. Additionally, as depicted in Figure \ref{fig:dconv}.a, the DConvB comprises two $3\times3$ convolutional blocks and a residual skip connection that uses a $1\times1$ convolutional kernel to match the input number of channels $C_{in}$ to $C_{out}$. The output of the two 3 by 3 kernels is summed with the features of the skip connection.

These pyramid feature maps play a crucial role in maintaining spatial attention awareness across all main encoder layers. They serve as gating signals for the main encoder path, facilitating the integration of spatial attention information throughout the encoding process.

\subsection{Main Encoder: Attention Fusion}

As shown in Figure \ref{fig:approach.2}, the input image is fed to both the Transformer and the main encoder branch. For the Transformer branch, we utilize PVT-v2-Li \cite{wang2022pvt}, which was designed for a progressive shrinking pyramid and a spatial-reduction attention. This makes the PVT flexible for learning multi-scale and high-level features, similar to the CNN encoder design. For the main branch, we start with a double convolution module as depicted in Figure \ref{fig:dconv}.a. From this point, it hierarchically merges the current features of the main branch and the Transformer features using a dual-gate attention mechanism (explained in the next section). The first level of the main branch is fused with the Transformer first stage features through the proposed Dual-Attention Gate. This attention fusion process is performed in the main branch for four levels. At each level the corresponding features of Transformer stage, pyramid levels, and previous main branch features are combined.

Upon the completion of the fourth fusion, the resulting features ($x_4$) are concatenated with the features from the previous level ($x_3$) in the main encoder. These are then fed into a classic ViT (ViT Base) with a spatial resolution of $14\times14$, corresponding to 196 tokens. The output features from the ViT ($x_6$) are subsequently passed through a dual convolution module. The resulting features ($x_7$) are then forwarded to the decoder. Finally, the decoder of the proposed PAG-TransYnet consists of four stages and follows a conventional architecture, with skips provided by two levels of features from the main encoder.



\subsection{Dual-Attention Gate}

The Dual-Attention gates play a crucial role in providing an effective fusion mechanism for the Transformer features at four different stages and the  features of the main encoder branch with attention provided by the pyramid level convolutional features. As depicted in the figure \ref{fig:dag}, this module has three inputs: Transformer features, the previous main branch level features, and the features associated with the corresponding pyramid level. The module consists of two classical Attention Gates (AG). The first AG considers the signal and the current features as input, while the second AG considers the current features as the signal and the lower pyramid feature as the gate. The outputs of both AGs are then concatenated to form the signal in the main Encoder, which is used in the skips to decoder levels. Each module incorporates Max pooling and Up sampling to match the spatial resolution of all three input features.

In summary, the main encoder branch receives the output of the convolution block at the input image of resolution $H\times W$, and then four Dual-Attention gates are utilized to obtain the encoded features, which is passed through DConvB to extract the next level features of the main branch. Both Dual-Attention Gate and DConvB are used to fuse then extract higher features, respectively, constructing a strong encoder for medical imaging segmentation.

\section{Datasets and tasks}

For abdominal organ segmentation, we utilized the Synapse multi-organ segmentation dataset, which has emerged as a benchmark dataset for evaluating the performance of medical imaging segmentation approaches in recent years. Following the precedent set by many state-of-the-art works, we adopted the training and validation splits introduced in the TransUnet paper \cite{chen2021transunet}.
In summary, the Synapse dataset consists of 30 abdominal CT scans introduced first hand in the MICCAI 2015 Multi-Atlas Abdomen Labeling Challenge and it has the pixel level annotation  of  8 abdominal organs (aorta, gallbladder, spleen, left kidney, right kidney, liver, pancreas, spleen, and stomach).

For infection segmentation tasks, we focused on the multi-class segmentation of Covid-19, specifically Ground Glass Opacity (GGO) and Consolidation, along with Bone Metastasis (BM) segmentation. These tasks present significant challenges due to the variability in infection shape, position, intensity, and type.
For Covid-19 segmentation, we followed the methodology outlined in \cite{bougourzi_emb-trattunet_2024,bougourzi2023d} and utilized two datasets from \cite{COVID-19-Dataset}. In total, we used 879 slices for training and 50 slices for testing. Among the training slices, 345 and 272 slices contained GGO and Consolidation infection types, respectively. The remaining slices without infection were included to enable the models to learn more features about healthy tissues.
For BM segmentation, we utilized the BM-Seg dataset \cite{afnouch2023bm}, employing the same data splits as in \cite{afnouch2023bm}. The BM-Seg dataset comprises 23 CT-scans, each covering one of multiple organs depending on the spread and primary cancer (e.g., lung, breast). In total, the dataset contains 1517 slices, and we employed a five-fold cross-validation strategy to evaluate the performance of the segmentation models.

For Gland and Nuclear segmentation tasks, we utilized two distinct datasets: the Gland segmentation dataset (GlaS) \cite{siri2017gland} and the MoNuSeg dataset \cite{kumar2019multi}, respectively. The GlaS dataset comprises 165 images specifically designed for gland segmentation tasks. On the other hand, the MoNuSeg dataset consists of 44 images tailored for nuclear segmentation tasks. Following the evaluation scheme proposed in \cite{wang2022uctransnet}, we conducted three iterations of five-fold cross-validation for each task. This approach ensures robust evaluation by splitting the dataset into five subsets, using each subset as a validation set once while training on the remaining four subsets. The results are corresponding to the  mean and standard deviation of the three runs, where each run result corresponds the five folds cross-validation results.

\begin{table}
 \caption{Comparison on Abdominal Multi-Organs Segmentation. DSC and HD95 are the average dice score and 95\% Hausdorff distance of the 8 classes, respectively. The fourth column to the last show the Dice-score (DSC) for each class.  }
 \begin{center}
\label{tab:synapse}
\centering
\resizebox{1\columnwidth}{!}{
\begin{tabular}{|l||c|c||c|c|c|c|c|c|c|c|}

\hline
 {\multirow{2}{*} \textbf{Architecture}}    & \multicolumn{2}{|c|}{\textbf{Average}}& \multirow{2}{*} \textbf{Aorta} &\multirow{2}{*} \textbf{Gallbladder} &\multirow{2}{*} \textbf{Kidney (L) } &\multirow{2}{*} \textbf{Kidney (R)}& \multirow{2}{*} \textbf{Liver}& \multirow{2}{*} \textbf{Pancreas}& \multirow{2}{*} \textbf{Spleen}& \multirow{2}{*} \textbf{Stomach}\\
       
\cline{2-3}
& \textbf{DSC$\uparrow$}& \textbf{HD95$\downarrow$} & &   & &  & &   &   &  \\
\hline\hline 

Unet \cite{oktay_attention_2018}&  74.68 &36.87& 84.18& 62.84 &79.19& 71.29& 93.35 &48.23 &84.41 &73.92 \\\hline

Att-Unet \cite{oktay_attention_2018}&75.57& 36.97& 55.92& 63.91& 79.20& 72.71 &93.56& 49.37& 87.19 &74.95 \\\hline
V-Net \cite{oktay_attention_2018}&  68.81& \bf{-} &75.34& 51.87 &77.10 &80.75 &87.84 &40.05 &80.56 &56.98 \\\hline 

 TransUnet \cite{chen2021transunet}& 77.48& 31.69 &87.23 &63.13 &81.87& 77.02 &94.08& 55.86 &85.08 &75.62  \\\hline

 MTUnet \cite{wang2022mixed}& 78.59 & 26.59 &  87.92&   64.99& 81.47& 77.29 & 93.06& 59.46  &  87.75 & 76.81\\\hline

 UCTransNet \cite{wang2022uctransnet}& 78.23 & 26.75 & \bf{-} &  \bf{-} & \bf{-}& \bf{-} & \bf{-}&  \bf{-} &  \bf{-} & \bf{-}\\\hline

TransClaw U-Net \cite{wang2022uctransnet}& 78.09 & 26.38 &85.87 & 61.38  & 84.83& 79.36 & 94.28&  57.65 &  87.74 &  73.55\\\hline
          
ST-Unet \cite{zhang2023st} &78.86 &20.37 &85.68& 69.05& 85.81&  73.04&  95.13&60.23&   89.15&  72.78
\\\hline

Swin-Unet \cite{liu_swin_2021} &77.58 &27.32 & 81.76 &65.95 &82.32 &79.22 &93.73& 53.81 &88.04 &75.79 \\\hline

VM-UNet \cite{ruan2024vm}& 81.08& 19.21& 86.40 &69.41& 86.16 &82.76& 94.17 &58.80 &89.51 &81.40 \\\hline

TransCeption \cite{azad2023enhancing}& 82.24& 20.89 &87.60 &\bf{71.82}& 86.23 &80.29 &95.01 &65.27 &91.68 &80.02
\\\hline

\bf{Ours} & \bf{83.43}&  \bf{15.82} &\bf{89.67} & 68.89 &\bf{86.74} & \bf{84.88}  &\bf{95.87} & \bf{68.75} &\bf{92.01} & \bf{80.66}   \\\hline
\end{tabular}}
\end{center}
\end{table}

\section{Experiments and Results}

\subsection{Experimental~Setup}
To produce our experiments, we mainly used PyTorch \cite{paszke_pytorch_2019} library for deep learning. Each architecture is trained for 100 epochs with an initial learning rate of 0.1 and Adam optimizer. The batch size is set to 16 images. The used machine has NVIDIA RTX A5000 GPU with 24 GB of memory, 11th Gen Intel(R) Core(TM) i9-11900KF (3.50GHz) CPU and 64 of RAM. Three types of active data augmentation are used; random rotate with an angle between $-35^{\circ}$ and $35^{\circ}$ with a probability of 10\% and  random Horizontal  and vertical Flipping with probability of 20\% for each.

\subsection{Results}
Tables \ref{tab:synapse}, \ref{tab:bm}, \ref{tab:covid}, and \ref{tab:micro} summarizes the comparison results with the state-of-the-art architectures in Synapse, BM-Seg, Covid-19, and GlaS and MoNuSeg datasets, respectively. These results show the superiority of our approach over the state-of-the-art architectures.

For the Synapse dataset results (Table \ref{tab:synapse}), we selected comparison approaches that followed the same evaluation splits as \cite{chen2021transunet}. Comparing with the TransUnet architecture, considered as the baseline for the Synapse dataset, our architecture demonstrated superior performance with improvements of 5.95\% and 15.87 for Dice-Score and HD95, respectively. This indicates the efficacy of our approach in leveraging both Transformer and CNN features through the proposed Dual-Attention Gate.

Furthermore, our architecture surpassed state-of-the-art methods in terms of both average Dice-Score and HD95 metrics, demonstrating superior performance across all segmented classes. However, it is noteworthy that our approach did not achieve the highest segmentation performance specifically for the Gallbladder class.

\begin{table}
\centering
\caption{Comparison on Bone Metastasis Segmentation. F1-S, DSC and IoU are F1-score, Dice-Score and Intersection over Union, respectively.}
\label{tab:bm}
\begin{tabular}{|l||c|c|c|}
\hline
\textbf{Model}         & \textbf{F1-S$\uparrow$} & \textbf{DSC$\uparrow$}     & \textbf{IoU$\uparrow$}          \\ \hline\hline
U-Net   \cite{afnouch2023bm}    & 79.46& 72.26 & 65.93  \\ \hline
AttUnet  \cite{afnouch2023bm}  & 79.41& 71.76& 65.86 \\ \hline
Unet++  \cite{afnouch2023bm}  & 79.74& 71.99& 66.31  \\ \hline
AttUnet++  \cite{afnouch2023bm} & 80.28 & 72.36 & 67.06  \\ \hline \hline

SwinUnet \cite{cao2022swin} &  61.09   &  39.17  &  44.01   \\\hline   
MTUnet \cite{wang2022mixed} &  58.59  &  44.30   &  41.45  \\\hline   
MISSFormer \cite{huang2022missformer}& 81.44 & 70.42 &  68.73 \\  \hline
UCTransNet \cite{wang2022uctransnet}& 83.62 &  73.88& 71.85  \\ \hline\hline

Hybrid-AttUnet++ \cite{afnouch2023bm}& 82.27 & 75.70 & 69.89  \\ \hline 
EDAUnet++ \cite{afnouch2023bm}&  83.67 & 77.05 &71.92    \\ \hline \hline

\textbf{Ours}  & \textbf{85.01} & \textbf{79.70} & \textbf{73.92}   \\ \hline

\end{tabular}

\end{table}


In the BM segmentation comparison, we present the results obtained by comparing our method with the competing approaches outlined in the dataset paper \cite{afnouch2023bm} and four recent transformer-based architectures: SwinUnet \cite{cao2022swin}, MTUnet \cite{wang2022mixed}, MISSFormer \cite{huang2022missformer}, and UCTransNet \cite{wang2022uctransnet}. Our proposed approach showcased superiority over these architectures (see Table \ref{tab:bm}).

Moreover, the lower performance of Transformer-based architectures, such as SwinUnet and MTUnet, raises concerns about their ability to generalize across different tasks, especially for infection segmentation tasks. Infection segmentation tasks involve high variability in shape, type, position, and intensity of infections, which may cover only a small portion compared to the background. 

In contrast, our approach exhibits a high ability to segment infection regions due to the rich features extracted and combined during the encoding phase. Additionally, the proposed Dual-Attention Gate effectively highlights prominent parts through multi-scale feature maps, making it well-suited for detecting infection regions.


Table \ref{tab:covid} provides a comprehensive summary of the results achieved by our proposed approach and its comparison with three baseline CNN architectures (U-Net, Att-Unet, and Unet++), four state-of-the-art approaches for Covid-19 segmentation (CopleNet \cite{wang_noise-robust_2020}, AnamNet \cite{paluru_anam-net_2021}, SCOATNET \cite{zhao2021scoat}, and EMB-TrAttUnet \cite{bougourzi_emb-trattunet_2024}), and four recent Transformer-based medical imaging segmentation approaches (SwinUnet \cite{cao2022swin}, MTUnet \cite{wang2022mixed}, MISSFormer \cite{huang2022missformer}, and UCTransNet \cite{wang2022uctransnet}).

Our analysis revealed that Transformer-based approaches exhibit limited generalization ability, achieving performance close to that of baseline CNN architectures. Additionally, a significant performance gap was observed between the segmentation of the two classes, primarily due to the minor presence of Consolidation compared to GGO, both in appearance and distribution within the lung.
Remarkably, our proposed approach achieved the best performance, effectively reducing the gap in segmenting both classes compared to the comparison approaches. This highlights our method's exceptional capability to accurately highlight infection regions throughout all encoding blocks, leveraging the proposed Dual-Attention Gates.


\begin{table}
 \caption{Comparison on Multi-classes Covid-19 Segmentation. F1-S, DSC and HD95 are F1-score, Dice-Score and 95\% Hausdorff distance, respectively. GGO and Con are the two types of Covid-19 infection known as Ground-Glass Opacity and Consolidation. }
 \begin{center}
\label{tab:covid}
\centering
\resizebox{\columnwidth}{!}{
\begin{tabular}{|l||c|c|c||c|c||c|c|}

\hline

 {\multirow{2}{*} \textbf{Architecture}}    & \multicolumn{3}{|c|}{\textbf{Average}}& \multicolumn{2}{|c|}{\textbf{F1-S}} &\multicolumn{2}{|c|}{\textbf{DSC}} \\

\cline{2-8}


& \textbf{F1-S$\uparrow$}& \textbf{DSC$\uparrow$}& \textbf{HD95$\downarrow$} & \textbf{GGO}& \textbf{Con} & \textbf{GGO}& \textbf{Con}\\
\hline\hline 

U-Net \cite{oktay_attention_2018}& 48.58& 32.79 &  35.69  & 
65.81$\pm$1.26 &31.35$\pm$12.96& 50.13$\pm$1.31& 15.45$\pm$5.66 \\\hline

Att-Unet \cite{oktay_attention_2018}& 51.92 & 34.85 &  35.84 & 64.81$\pm$1.89 &
39.04$\pm$6.81  & 50.44$\pm$1.35& 19.26$\pm$3.55 \\\hline
Unet++ \cite{zhou_unet_2018}&48.51 & 41.48 & 44.06 & 
65.69$\pm$1.29&
31.31$\pm$6.67 &51.65$\pm$4.12 & 31.31$\pm$6.67\\\hline\hline 

CopleNet \cite{wang_noise-robust_2020}& 54.64 &31.355 & 39.04& 60.44$\pm$1.54  &  29.70$\pm$10.29 &   46.25$\pm$3.13& 16.46$\pm$4.76 \\\hline

AnamNet \cite{paluru_anam-net_2021}& 48.53&  34.875 &  34.78 &65.10$\pm$ 3.56  &  31.97$\pm$6.12    &51.69$\pm$4.8 & 18.06$\pm$4.61  \\\hline

SCOATNET  \cite{zhao2021scoat}& 45.07& 37.06& 30.99 & 65.77$\pm$3.28 & 43.52$\pm$1.67   &   50.80$\pm$4.63 & 23.32$\pm$2.07    \\\hline\hline

 SwinUnet \cite{cao2022swin}& 47.47 & 31.11 &39.42 &62.74$\pm$2.63
&32.2$\pm$6.68&42.46$\pm$2.61&
19.77$\pm$3.87 \\\hline
          
MTUnet \cite{wang2022mixed}& 42.30 & 30.60& 37.50 &57.83$\pm$2.57&
26.78$\pm$7.39&
42.97$\pm$2.78&18.24$\pm$ 4.56  \\\hline

MISSFormer \cite{huang2022missformer} & 56.70 & 39.79 & 42.08 &  65.66 $\pm$3.06&
47.75$\pm$4.77&51.57$\pm$4.01&
28.02$\pm$2.72\\\hline
 
UCTransNet \cite{wang2022uctransnet}&58.33 &  41.41 & 34.67& 67.46$\pm$2.97&
49.21$\pm$4.27 &   53.42$\pm$4.24 &29.41$\pm$3.48  \\\hline\hline

EMB-TrAttUnet \cite{bougourzi_emb-trattunet_2024}& 65.16&  48.18 & 27.47 & 70.06$\pm$0.03& 60.26$\pm$0.92 &  59.14$\pm$0.87  & 37.23$\pm$0.97 \\\hline\hline

\bf{Ours}& \bf{68.71}&  \bf{51.03} & \bf{24.22}  &   \bf{73.12$\pm$0.37}&
\bf{64.30$\pm$0.90}  & \bf{60.38$\pm$0.94} & \bf{41.68$\pm$0.98}  \\\hline

\end{tabular}}
\end{center}
\end{table}
\begin{table}
 \caption{Comparison on Glas and  MoNuSeg Segmentation datasets.   }
 \begin{center}
\label{tab:micro}
\centering
\begin{tabular}{|c|l|c|c||c|c|}

\hline
 Ex &{\multirow{2}{*}    \textbf{Architecture}}    & \multicolumn{2}{|c|}{\textbf{GlaS}}& \multicolumn{2}{|c|}{\textbf{MoNuSeg}} \\
\cline{3-6}


&  &\textbf{DSC} &   \textbf{IoU} &\textbf{DSC} &   \textbf{IoU}  \\
\hline

1& U-Net \cite{wang2022uctransnet}& 85.45$\pm$1.3  &  74.78$\pm$1.7  & 76.45$\pm$2.6 & 62.86$\pm$3.0 \\\hline 

2& Unet++ \cite{wang2022uctransnet}& 87.56$\pm$1.2  &  79.13$\pm$1.7  & 77.01$\pm$2.1 & 63.04$\pm$2.5\\\hline 

3& AttUNet \cite{wang2022uctransnet}& 88.80$\pm$1.1  &  80.69$\pm$1.7  & 76.67$\pm$1.1 & 63.47$\pm$1.2 \\\hline 

4& MRUNet \cite{wang2022uctransnet}& 88.73$\pm$1.2  &  80.89$\pm$1.7  & 78.22$\pm$2.5 & 64.83$\pm$2.9 \\\hline

5& TransUNet \cite{wang2022uctransnet}& 88.40$\pm$0.7  &  80.40$\pm$1.0  & 78.53$\pm$1.1 & 65.05$\pm$1.3 \\\hline

6& MedT \cite{wang2022uctransnet}& 85.92$\pm$2.9  &  75.47$\pm$3.5  & 77.46$\pm$2.4 & 63.37$\pm$3.1 \\\hline

7& Swin-Unet \cite{wang2022uctransnet}& 89.58$\pm$0.6  &  82.06$\pm$ 0.7 & 77.69$\pm$0.9 & 63.77$\pm$ 1.2\\\hline

8& UCTransNet \cite{wang2022uctransnet}& 90.18$\pm$0.7  &  82.96$\pm$1.1  & 79.08$\pm$0.7 & 65.50$\pm$0.9 \\\hline


9&\bf{Ours}& \bf{94.20$\pm$0.55} & \bf{89.29$\pm$0.91}& \bf{79.62$\pm$0.7}& \bf{66.31$\pm$0.6} \\\hline 
\end{tabular}
\end{center}
\end{table}

\begin{table}
\caption{Ablation study on Synapse Dataset and Covid-19. The importance of the following elements is studied: CNN Pyramid path (Pyr), PVT path (PVT) and the Vit Transformer (ViT). Mean Dice-Score (DSC) and 95\% Hausdorff distance metrics are used for both tasks plus F1-Score (F1-S) for for Covid-19 task.}\label{tab:abl}
 \begin{center}
\centering
\scalebox{1}{\begin{tabular}{|l||ccc||c|c||c|c|c|}

\hline
  {\multirow{2}{*}    \textbf{Architecture}}   &\multicolumn{3}{|c|}{\textbf{Ablation}}  & \multicolumn{2}{|c|}{\textbf{Synapse}}& \multicolumn{3}{|c|}{\textbf{Covid-19}}  \\

\cline{2-9}


 &\textbf{Pyr} &\textbf{PVT}&\textbf{ViT}   &\textbf{DSC$\uparrow$} &   \textbf{HD95$\downarrow$}& \textbf{F1-S$\uparrow$} &\textbf{DSC$\uparrow$} &   \textbf{HD95$\downarrow$} \\
\hline


(1) No Pyramid Path&\xmark& \cmark & \cmark & 82.32 & 21.45 & 67.84 & 51.07 & 23.23\\\hline

(2) No PVT &\cmark& \xmark & \cmark&79.44 &  22.92  & 65.98 & 50.25 &24.05\\\hline

(3) No ViT&\cmark& \cmark &  \xmark & 82.39& 17.67 & \bf{68.92} &\bf{51.69} & \bf{21.53}\\\hline

\bf{(4) PAG-TransYnet} &\cmark& \cmark &  \cmark& \bf{83.43} & \bf{15.82}& 68.71& 51.03& 24.22  \\\hline


\end{tabular}}
\end{center}
\end{table}


Following the evaluation protocol and comparing the performance with the results obtained in \cite{wang2022uctransnet}, Table \ref{tab:micro} presents a comprehensive comparison of our approach with the state-of-the-art methods for microscopic segmentation tasks, specifically Gland and Nucleus segmentation.
From these results, it is evident that our proposed architecture outperforms the state-of-the-art methods, achieving the best performance in both Gland and Nucleus segmentation tasks. This further confirms the efficiency and versatility of our approach in various medical imaging segmentation tasks.

\subsection{Ablation Study}  

The aim of this section is to investigate the significance of the proposed encoding elements within our approach. We examine the importance of the following components: CNN Pyramid path (Pyr), PVT path (PVT), and the ViT Transformer (ViT), considering multi-organ abdominal segmentation (Synapse) and 
 infection segmentation (Covid-19). The results are summarized in Table \ref{tab:abl}.
In the first ablation experiment, it is evident that the Pyramid path plays a crucial role in Synapse segmentation, as removing it leads to a decrease in performance by 1.11\% and 5.63 for DSC and HD95, respectively. Conversely, the results for Covid-19 segmentation show stable performance despite removing the Pyramid path.

In the second ablation study, it becomes apparent that Transformer features are vital for both tasks. Removing the PVT path results in a significant decrease in performance on the Synapse dataset, with a reduction of 4\% and 7.1 for Dice-score and HD95, respectively. Similarly, for Covid-19 segmentation, the performance decreases by 2.73\% and 0.78\% for F1-score and Dice-score, respectively.
Regarding the ViT block, the experiments demonstrate its importance in Synapse segmentation, likely due to the complexity of Synapse having more classes compared to Covid-19 segmentation. Additionally, the relatively smaller size of the Covid-19 dataset makes it challenging to train the ViT (base varaint), leading to potential overfitting. However, the experiments show only a minor decrease in performance.  Overall, these findings underscore the significance of each component in achieving high performance in both Synapse and Covid-19 segmentation tasks, with particular emphasis on the Transformer features in enhancing segmentation accuracy.


\section{Conclusion}

In this paper, we introduce a novel hybrid architecture, termed PAG-TransYnet, designed for medical imaging segmentation. By seamlessly integrating Convolutional Neural Networks (CNNs), Transformers, and a fusion branch encoder, we aim to address the limitations of existing approaches and improve segmentation accuracy. Our key innovation lies in enhancing the Att-Unet attention gate with our proposed Dual-Attention Gate mechanism. This mechanism facilitates the extraction of prominent features from multiple encoder branches, thereby capturing both local and global contextual information more effectively.

Through comprehensive evaluation across various segmentation tasks, including abdominal multi-organs segmentation, infection detection (Covid-19 and Bone Metastasis), and microscopic tissue segmentation (Gland and Nucleus), our proposed approach demonstrates state-of-the-art performance and remarkable generalization capabilities. The utilization of the Dual-Attention Gate mechanism enables efficient fusion of features from different encoder branches, leading to enhanced segmentation accuracy and robustness across diverse medical imaging datasets.

The contributions of this work extend beyond the development of a novel segmentation architecture. We present a significant advancement towards addressing the pressing need for efficient and adaptable segmentation solutions in medical imaging applications. By seamlessly integrating CNNs and Transformers, our approach provides a versatile framework capable of handling the high variability among diseases and segmentation tasks. Furthermore, our methodology lays the foundation for future research endeavors aimed at advancing medical imaging segmentation techniques and facilitating clinical decision-making processes.



%
%
%


\end{document}